%% file: script.tex
\documentclass[doublecol]{epl2}
\usepackage{epsfig}

\title{$N$-body Efimov states {of} trapped {Bosons}}

\author{M.~Th{\o}gersen, D.V.~Fedorov, and A.S.~Jensen}

\institute{IFA, Aarhus University, 8000 Aarhus C, Denmark}


\abstract{
We demonstrate the possibility of existence of meta-stable $N$-body Efimov
states in trapped Bose {systems} with large scattering length.
We calculate spectra of trapped systems of $N$=3,4,5,6, and 7 bosons
using a stochastic variational method with a restricted correlated
Gaussian basis. For each system the calculations reveal a series
of Efimov states where the energy and the r.m.s. radius exhibit the
characteristic exponential dependence upon the state number. {We
also estimate the contribution of these states to the recombination rate
of Bose-Einstein condensates.}}

\begin{document}

\maketitle

\section{Introduction}

The Efimov effect~\cite{efimov} appears in quantum three-body systems
when attractive interactions between at least two pairs of particles
are such that the scattering length is much larger than the range of
the interaction; in other words two of the three two-body subsystems are
close to the threshold of binding. Under these conditions a characteristic
series of weakly bound and spatially extended states, called Efimov
states~\cite{fedorov93}, appears in the system.  These states appear due
to specific long range two-body correlations between particles caused
by the large scattering length.

The effect is easiest to see in the hyper-spherical adiabatic
approximation where the slow adiabatic variable is the hyper-radius $\rho$
(root square radius of the system). It has been shown~\cite{fedorov93}
that close to the two-body threshold the effective adiabatic potential
$W(\rho)$ is attractive and asymptotically proportional to the inverse
square of the hyper-radius,
\begin{equation}
W(\rho)=-{\hbar^2\over2m}{\xi^2-{1\over4}\over\rho^2}\;,
\end{equation}
where $m$ is the mass scale, and $\xi$ is a constant depending on masses
of the particles~\cite{nielsen,nielsen98}.

A sufficiently large positive $\xi^2$ leads to a geometric series of bound
states with exceedingly small energies, $E_n\propto e^{-\zeta\cdot n}$, and
exceedingly large root mean square radii, $R_n\propto
e^{{1\over2}\zeta\cdot n}$, where $n$ is the state number and
$\zeta={2\pi\over\xi}$.

There is a general theoretical consensus~\cite{lim77,esry96,nielsen98}
that the first excited state of the helium trimer $^4$He$_3$ is an
Efimov state, although the experimental observation so far proved
elusive~\cite{tri-exp-05}.

The three-body Efimov effect has recently got some
support~\cite{kraemer06} from an experiment with trapped Bose gases
where the recombination rate was measured as function of the scattering
length. The latter was varied using the Feschbach resonance technique by
applying an external magnetic field. A sharp peak in the recombination
rate was detected and interpreted as a three-body Efimov resonance in
qualitative agreement with theoretical predictions~\cite{esry99}.

It has been shown that in an $N$-body system with $N>$3 the Efimov effect
does not exist at the $N-1$ threshold~\cite{amado73}.  At the two-body
threshold the $N$-body Efimov states with $N>3$ can not exist either as
the clusters with 3 and higher number of particles are generally deeply
bound at this point. The Efimov states would then be unstable due to
lower lying thresholds and would decay into deeply bound cluster states.

However it has been suggested in~\cite{ole02} that a sequence
of meta-stable $N$-body states with the characteristic exponential
energy dependence can yet show up at the two-body threshold.  Using the
$N$-body hyper-spheric method it has been shown that an $N$-body system
at the two-body threshold has a hyper-spheric adiabatic potential with
inverse-square dependence. This peculiar adiabatic potential appears
due to the same mechanism as for three particles and thus gives rise to
$N$-body Efimov states with a structure similar to that of three-body
Efimov states: an (otherwise) uncorrelated system with very specific
two-body correlations caused by the large scattering lengths.

This specific hyper-spheric adiabatic potential is not the lowest
one as different bound clusters with lower thresholds create lower
lying adiabatic potentials.  However, although not truly bound, these
$N$-body Efimov states might still exist as meta-stable states slowly
decaying into clusters, much like the Bose-Einstein condensate states.
The structure of the Efimov states is determined by the long-range
two-body correlations and is thus quite dissimilar to the structure of
clusterized states with short-range many-body correlations. Therefore
the overlap between Efimov states and clusterized states should be small
and consequently the life-time should be large.

The conclusions about meta-stable $N$-body Efimov states were obtained
in~\cite{ole02} in an extreme hyper-spheric adiabatic approximation
where the couplings to all other channels were neglected.  In this
letter we report on a more realistic calculation of $N$-body
Efimov states with a different method, namely the restricted correlated
Gaussian stochastic variational method where no adiabatic approximation
is assumed.

\section{The system and the method}

We consider a system of $N$ identical bosons with mass $m$ and coordinates
$\mathbf{r}_{i}$ in a spherical harmonic trap with frequency $\omega$,
{where the scattering length is assumed to be tuned to a large
value, facilitating the Efimov effect}.

The Hamiltonian of the system is given by
\begin{equation}\label{eq:ham}
H=-\frac{\hbar^{2}}{2m}
\sum_{i=1}^{N}\frac{\partial^2}{\partial\mathbf{r}_{i}^{2}}
+\sum_{i<j}V(\left|\mathbf{r}_{i}-\mathbf{r}_{j}\right|)
+\frac{m\omega^{2}}{2}\sum_{i=1}^{N}r_{i}^{2}\;,\label{eq:h}
\end{equation}
where the system parameters are $m=86.909$~u, and the trap length
$b_t=\sqrt{\hbar/(m\omega)}$=46189~au.

The two-body potential is an attractive Gaussian with the range
$b$=11.65au and depth 1.24825au corresponding to the scattering length
of $a$=$-1.4\cdot10^6$~au~$\gg b_t$.

The wave-function of the system is represented as a linear combination of
$K$ basis-functions taken in the form of symmetrized correlated
Gaussians,
\begin{equation}
\Psi=
\hat{S}\sum_{k=1}^{K}C_{k} \exp
\left(
-\frac{1}{2}\sum_{i<j}^{N}\alpha_{ij}^{(k)}
(\mathbf{r}_{i}-\mathbf{r}_{j})^2
\right)
\;,\label{eq:psi-full}
\end{equation}
where the total angular momentum is zero, $\hat{S}$ is the
symmetrization operator, and $C_{k}$ and $\alpha_{ij}^{(k)}$ are
variational parameters. The linear parameters $C_{k}$ are determined
by diagonalization of the Hamiltonian while the non-linear parameters
$\alpha_{ij}^{(k)}$ are optimized stochastically~\cite{varga,hansh} by
random sampling from a region that covers the spatial distances from $b$
to $b_t$.

During calculations of a given system the number of Gaussians in the
basis is increased and the stochastic optimization is carried out until
all energy levels of interest are converged.

{With uncorrelated Gaussians the method is
equivalent~\cite{hansh} to a mean-field approximation which is unable
to describe the Efimov effect and which has the validity condition
$na^3\ll1$. Introducing successive correlations beyond the mean-field
improves this validity condition and allows calculations of the Efimov
states.}

{The structure of the} $N$-body Efimov states {is}
analogous to that of the three-body Efimov states:  the spatial extension
of {these states is} much larger than the range of the potential
$b$, hence the density of the system is small, $nb^{3}\ll1$; and there
are no cluster substructures. {For low-density systems,
$nb^{3}\ll1$, only two-body correlations are of importance.}

Allowing only two-body correlations,
the variational wave-function can be simplified as
\begin{eqnarray}
\Psi_{2b} &=& \hat{S}\sum_{k=1}^{K}C_{k} \nonumber \\ &\times&\exp\left(
-\frac{1}{2}\alpha^{(k)}\rho^{2} -\frac{1}{2}\beta^{(k)}({\mathbf
r}_{1}-{\mathbf r}_{2})^{2}\right) \;, \label{eq:psi-2b}
\end{eqnarray}
where $\rho^2=\sum_{i<j}(\mathbf{r}_i-\mathbf{r}_j)^2$ is the hyper-radius
and $\alpha^{(k)}$ and $\beta^{(k)}$ are the nonlinear parameters. The
symmetrization of this function can be done analytically \cite{hansh}
which greatly simplifies the numerical calculations.

This form of the variational space proved very successful in describing
the Bose-Einstein condensate states~\cite{martin07} which have
similar structure. The clusterized states, uninteresting in the present
context, are explicitly excluded while the crucial two-body correlations
responsible for the existence of the Efimov states are retained. Otherwise
it would be impossible to calculate the Efimov states due to the enormous
number of cluster thresholds and clusterized states in an $N$-body
system. {Omitting the} clusterized states, {into which
the Efimov states would decay}, {effectively disregards the
widths of the Efimov states. However the clusterized states} differ
{significantly} in spatial structure and therefore should overlap
very little with the computed {Efimov} states.  {Indeed
an estimate~\cite{penkov} of the lifetimes of the molecular three-body
Efimov states showed that their widths are extremely small and that they
can be considered bound for all practical purposes.}

\section{Results}

\begin{figure}
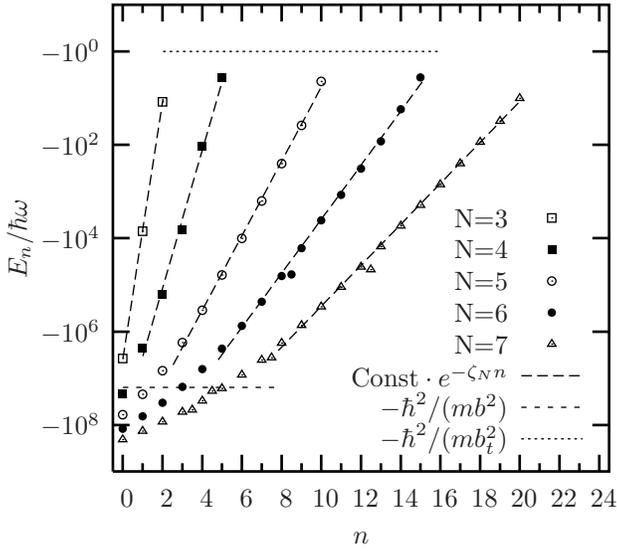

\centerline{\input fig-e.tex}
\caption{Energy $E_n$ as function of the state number $n$ for negative
energy states of a system of $N$ bosons in a harmonic trap with length
$b_t$ interacting via an attractive potential of range $b$ with scattering
length much larger than $b_t$. The horizontal lines demarcate the region
where the Efimov states can exist in trapped gases. The lines show the
exponential fits of the form $E_n\propto e^{-\zeta\cdot n}$ drawn through the
points in the indicated region.}

\label{fig:e}\end{figure}

The energies of Efimov states in a trap should be on one hand much
larger than the typical energy scale of a cluster state, ${\hbar^2
\over mb^2}$, and on the other hand smaller than the oscillator energy
$\hbar\omega={\hbar^2\over mb_t^2}$.

We have calculated the spectrum of trapped boson systems with the
Hamiltonian eq.~(\ref{eq:ham}) with $N$=3,4,5,6, and~7. The calculated
energies are shown on Fig.~\ref{fig:e}. Indeed in the indicated energy
region for each of the $N$-body systems there is a series of states with
exponential dependence upon the state number, $E_n\propto e^{-\zeta_N n}$.

\begin{figure}
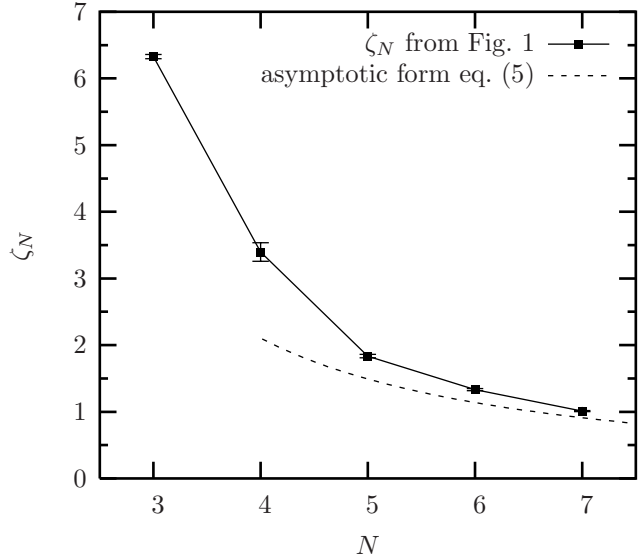

\centerline{\input fig-z.tex}
\caption{The exponents $\zeta_N$ from Fig.\ref{fig:e} as function of the
boson number $N$ together with the asymptotic
form eq.~(\ref{eq:zN}).
}
\label{fig:z}\end{figure}

The exponential fits give the numbers $\zeta_3$=6.33$\pm0.032$,
$\zeta_4$=3.40$\pm0.14$, $\zeta_5$=1.79$\pm0.034$,
$\zeta_6$=1.31$\pm0.020$, and $\zeta_7$=1.01$\pm0.007$. The value for
$N$=3 agrees within 3 sigma with the known analytical result of 6.244
(see \cite{nielsen} and references therein). Fig.~\ref{fig:z} shows our
numerical results together with the asymptotic expression for large $N$,
\begin{equation}\label{eq:zN}
\zeta_N=\frac{2\pi}{\sqrt{
{5\over3}N^{7\over3}\left(1-{2\over N}\right)-{(3N-4)(3N-6)\over4}-{1\over4}
}}\;,
\end{equation}
which was established in~\cite{sogo}. Our results seem consistent
with the asymptotic estimate.

\begin{figure}
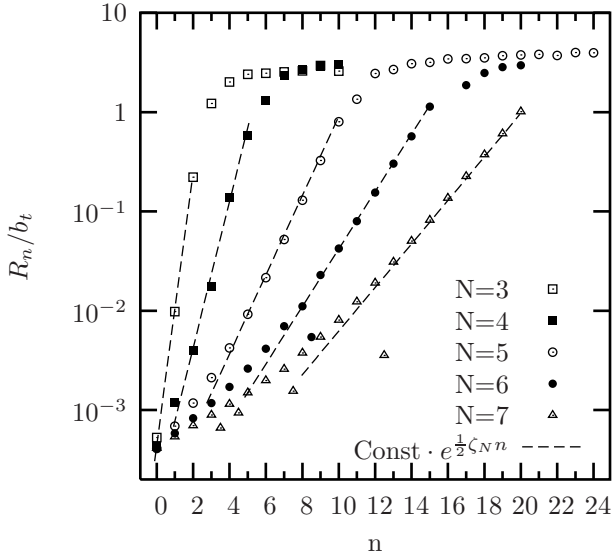

\centerline{\input fig-r.tex}
\caption{Root mean square radius $R_n$ as function of the state number $n$
of a system of $N$ bosons from Fig.\ref{fig:e}.  The lines show the fitting
curves of the form $R_n\propto e^{{1\over2}\zeta_Nn}$ drawn through the
same points and with the same parameters $\zeta_N$ as on Fig.\ref{fig:e}.
}
\label{fig:r}\end{figure}

The spatial extension of Efimov states in trapped systems must be
much larger than the interaction range $b$ and much smaller than
the trap length $b_t$.  On Fig.~\ref{fig:r} are shown the calculated
r.m.s. radii $R_n$ as function of state number $n$.  The radii of the
Efimov states identified on Fig.~\ref{fig:e} are reproduced well with
the exponentials $R_n\propto e^{{1\over2}\zeta_Nn}$ where the parameters
$\zeta_N$ are taken from the fits on Fig.~\ref{fig:e}.  Apparently all
these states fall within the correct boundaries and the values of the
radii follow the correct exponential trend.

There are several states in the $N$=6 and 7 systems with radii much
smaller than those of the typical states in the series with similar
energies. Clearly these states are not Efimov states but rather relatively
compact states with a different structure.

\section{Recombination reactions into shallow $N$-body states}

The $N$-body Efimov states could in principle be identified by their
contribution to the recombination rate of a cold gas as function of
scattering length similar to the three-body case~\cite{kraemer06}. As
the scattering length is increased using the Feschbach technique, the
$N$-body Efimov states crossing the threshold should produce peaks in
the recombination rate. Since the $n$-th Efimov states appears when the
scattering length is about $a\approx a_0e^{\frac{1}{2}\zeta\cdot n}$
(where $a_0$ is the scattering length corresponding to the lowest Efimov
state) the peaks will appear as an exponential sequence as function of
the scattering length~\cite{nielsenmacek}.

The   scale of the rate can be estimated as follows.  Let us first
consider a three-body reaction. The rate of the loss of particles
from a cold gas due to the three-body recombination reaction into
a shallow dimer with the energy $\frac{\hbar^{2}}{ma^{2}}$
is given by Fermi's golden rule,
\begin{equation}
-\frac{dN_{0}}{dt}=
3\frac{N_{0}^{3}}{6}\frac{2\pi}{\hbar}\left|T_{fi}\right|^{2}
\frac{d\nu_{f}}{dE_{f}}\;,\label{eq:dndt}
\end{equation}
where the factor 3 is there since each recombination reaction
removes three particles from the cold gas, $N_{0}^{3}/6$ is the
number of triples in the gas of $N_{0}$ particles, $d\nu_{f}$
is the number of final states (dimer plus the third particle) with
relative momentum $q_{f}=\frac{2}{\sqrt{3}}a^{-1}$ and the relative
kinetic energy $E_{f}\equiv\frac{\hbar^{2}q_{f}^{2}}{2(\frac{2}{3}m)}=
\frac{\hbar^{2}}{ma^{2}}$,
\begin{equation}
d\nu_{f}=\frac{Vd^{3}q_{f}}{(2\pi)^{3}}
=\frac{2}{3\sqrt{3}\pi{}^{2}}\frac{Vm}{\hbar^{2}a}dE_{f}\;,
\end{equation}
and $T_{fi}$ is the transition matrix element from the initial three-body
to the final dimer+particle state.

In the typical experimental regime, where the scattering length is
still much smaller than the size of the trap, $a\ll b_{t}$, the transition
matrix element for the non-resonant three-body recombination rate
from a cold gas state into a shallow dimer state can be estimated
perturbatively substituting the asymptotic expressions for the initial
and final wave-functions,
\begin{eqnarray}
&T_{fi}=\int d^{3}rd^{3}R
\left[
\psi_{\mathrm{d}}(r)\frac{e^{i\mathbf{q}_{f}\mathbf{R}}}{\sqrt{V}}
\right]&\nonumber\\
&\times\left(
U(\mathbf{R}-\frac{1}{2}\mathbf{r})
+U(\mathbf{R}+\frac{1}{2}\mathbf{r})
\right)
\left[
\frac{e^{i\mathbf{k}\mathbf{r}}}{\sqrt{V}}
\frac{e^{i\mathbf{q}\mathbf{R}}}{\sqrt{V}}
\right]&\;,
\label{eq:tfi1}
\end{eqnarray}
where $\mathbf{r}$ is the distance between two particles, $\mathbf{R}$
is the distance between their center of mass and the third particle,
$V\propto b_{t}^{3}$ is the normalization volume, $k$ and $q$
($k\sim q\propto b_{t}^{-1}\ll a^{-1}$) are the initial momenta of
the cold gas particles, $\psi_{\mathrm{d}}(r)$ is the $s$-wave function
of the shallow dimer with the binding energy $\frac{\hbar^{2}}{ma^{2}}$.
Using the zero-range approximation
for the transition interaction,
$U(\mathbf{r})=\frac{4\pi\hbar^{2}a}{m}\delta(\mathbf{r})$,
and for the dimer wave-function,
$\psi_{\mathrm{d}}(r)\propto a^{-1/2}\frac{e^{-\frac{r}{a}}}{r}$,
the matrix element (\ref{eq:tfi1}) in the limit $k\sim q\ll a^{-1}$
is estimated as (cf.~\cite{fedichev})
\begin{equation}
T_{fi}\propto\frac{\hbar^{2}a^{5/2}}{V^{3/2}m}\;.
\end{equation}

Finally, the non-resonant factor of the three-body recombination
rate becomes (cf.~\cite{nielsen,fedichev})
\begin{equation}
\left.-\frac{dn}{dt}\right|_{\mathrm{3-body}}\propto n^{3}\frac{\hbar
a^{4}}{m}\;,\end{equation}
where $n=N_{0}/V$ is the density.

For an $N$-body recombination reaction into a shallow $N$-1
body Efimov state with the binding energy of the order of
$\frac{\hbar^{2}}{ma^{2}}$ the modifications to the expression for the
rate (\ref{eq:dndt}) include the extra $3(N-3)$ spatial dimensions
in the integral in eq.(\ref{eq:tfi1}), which gives an extra factor
$(a^{3/2}V^{-1/2})^{2(N-3)}$, and also the factor $N_{0}^{3}/6$ is
substituted by $N_{0}^{N}/N!$. Thus for the $N$-body recombination rate we
have \begin{equation} \left.-\frac{dn}{dt}\right|_{\mathrm{N-body}}\propto
n^{3}\frac{\hbar a^{4}}{m}(na^{3})^{N-3}\;.  \end{equation}

We emphasize that these simple estimates only refer to the non-resonant
contributions and they also do not include other types of decays where
the final state structure might be substantially different from the
shallow $N$-1 cluster plus one particle.

Although in the regime $na^{3}\ll1$ the $N$-body recombination has an
additional small factor $(na^{3})^{N-3}$ it might still be possible to
observe the $N$-body Efimov states as a sequence of resonant peaks in the
recombination rate as function of scattering length. The ratio of
scattering lengths corresponding to the adjacent
peaks, say number $n+1$ and number $n$, is given as
$\frac{a^{(n+1)}}{a^{(n)}}=e^{{1\over2}\zeta_{N}}$.

\section{Conclusion}

We have calculated the spectrum of trapped $N$-boson systems with
$N$=3,4,5,6, and 7 using the stochastic variation method with a
restricted correlated Gaussian basis. Only two-body correlations were
allowed in the variational space.  Thus the cluster states were a priori
mostly excluded from the variation space, which made the calculations
technically possible.

For each system a series of states is found with specific exponential
dependences of the energies and r.m.s radii on the state number, which
is a characteristic feature of Efimov states. For the $N$=3 system the
exponent obtained agrees well with the known analytical value and the
trend also agrees well with the existing asymptotic estimate.

Inclusion of the cluster states would turn the Efimov states into
meta-stable states. However the life-time of these states should be
comparable with the Bose-Einstein condensate states which have a similar
structure and similar decay modes.

It might be possible to observe the 4-body Efimov states as peaks in the
recombination rate of a condensate as function of the scattering length,
similar to the 3-body case. However, for a dilute gas the 4-body
recombination rate is a factor $na^3$ smaller than the 3-body rate,
therefore the accuracy constraints on the experiment would be higher.

In conclusion we have lent theoretical support to the possibility of
existence of long lived meta-stable $N$-body Efimov states in trapped
Bose gases.

\end{document}

%% file: fig-e.tex
\begingroup%
  \makeatletter%
  \newcommand{\GNUPLOTspecial}{%
    \@sanitize\catcode`\%=14\relax\special}%
  \setlength{\unitlength}{0.1bp}%
{\GNUPLOTspecial{!
/gnudict 256 dict def
gnudict begin
/Color false def
/Solid false def
/gnulinewidth 5.000 def
/userlinewidth gnulinewidth def
/vshift -33 def
/dl {10 mul} def
/hpt_ 31.5 def
/vpt_ 31.5 def
/hpt hpt_ def
/vpt vpt_ def
/M {moveto} bind def
/L {lineto} bind def
/R {rmoveto} bind def
/V {rlineto} bind def
/vpt2 vpt 2 mul def
/hpt2 hpt 2 mul def
/Lshow { currentpoint stroke M
  0 vshift R show } def
/Rshow { currentpoint stroke M
  dup stringwidth pop neg vshift R show } def
/Cshow { currentpoint stroke M
  dup stringwidth pop -2 div vshift R show } def
/UP { dup vpt_ mul /vpt exch def hpt_ mul /hpt exch def
  /hpt2 hpt 2 mul def /vpt2 vpt 2 mul def } def
/DL { Color {setrgbcolor Solid {pop []} if 0 setdash }
 {pop pop pop Solid {pop []} if 0 setdash} ifelse } def
/BL { stroke userlinewidth 2 mul setlinewidth } def
/AL { stroke userlinewidth 2 div setlinewidth } def
/UL { dup gnulinewidth mul /userlinewidth exch def
      dup 1 lt {pop 1} if 10 mul /udl exch def } def
/PL { stroke userlinewidth setlinewidth } def
/LTb { BL [] 0 0 0 DL } def
/LTa { AL [1 udl mul 2 udl mul] 0 setdash 0 0 0 setrgbcolor } def
/LT0 { PL [] 1 0 0 DL } def
/LT1 { PL [4 dl 2 dl] 0 1 0 DL } def
/LT2 { PL [2 dl 3 dl] 0 0 1 DL } def
/LT3 { PL [1 dl 1.5 dl] 1 0 1 DL } def
/LT4 { PL [5 dl 2 dl 1 dl 2 dl] 0 1 1 DL } def
/LT5 { PL [4 dl 3 dl 1 dl 3 dl] 1 1 0 DL } def
/LT6 { PL [2 dl 2 dl 2 dl 4 dl] 0 0 0 DL } def
/LT7 { PL [2 dl 2 dl 2 dl 2 dl 2 dl 4 dl] 1 0.3 0 DL } def
/LT8 { PL [2 dl 2 dl 2 dl 2 dl 2 dl 2 dl 2 dl 4 dl] 0.5 0.5 0.5 DL } def
/Pnt { stroke [] 0 setdash
   gsave 1 setlinecap M 0 0 V stroke grestore } def
/Dia { stroke [] 0 setdash 2 copy vpt add M
  hpt neg vpt neg V hpt vpt neg V
  hpt vpt V hpt neg vpt V closepath stroke
  Pnt } def
/Pls { stroke [] 0 setdash vpt sub M 0 vpt2 V
  currentpoint stroke M
  hpt neg vpt neg R hpt2 0 V stroke
  } def
/Box { stroke [] 0 setdash 2 copy exch hpt sub exch vpt add M
  0 vpt2 neg V hpt2 0 V 0 vpt2 V
  hpt2 neg 0 V closepath stroke
  Pnt } def
/Crs { stroke [] 0 setdash exch hpt sub exch vpt add M
  hpt2 vpt2 neg V currentpoint stroke M
  hpt2 neg 0 R hpt2 vpt2 V stroke } def
/TriU { stroke [] 0 setdash 2 copy vpt 1.12 mul add M
  hpt neg vpt -1.62 mul V
  hpt 2 mul 0 V
  hpt neg vpt 1.62 mul V closepath stroke
  Pnt  } def
/Star { 2 copy Pls Crs } def
/BoxF { stroke [] 0 setdash exch hpt sub exch vpt add M
  0 vpt2 neg V  hpt2 0 V  0 vpt2 V
  hpt2 neg 0 V  closepath fill } def
/TriUF { stroke [] 0 setdash vpt 1.12 mul add M
  hpt neg vpt -1.62 mul V
  hpt 2 mul 0 V
  hpt neg vpt 1.62 mul V closepath fill } def
/TriD { stroke [] 0 setdash 2 copy vpt 1.12 mul sub M
  hpt neg vpt 1.62 mul V
  hpt 2 mul 0 V
  hpt neg vpt -1.62 mul V closepath stroke
  Pnt  } def
/TriDF { stroke [] 0 setdash vpt 1.12 mul sub M
  hpt neg vpt 1.62 mul V
  hpt 2 mul 0 V
  hpt neg vpt -1.62 mul V closepath fill} def
/DiaF { stroke [] 0 setdash vpt add M
  hpt neg vpt neg V hpt vpt neg V
  hpt vpt V hpt neg vpt V closepath fill } def
/Pent { stroke [] 0 setdash 2 copy gsave
  translate 0 hpt M 4 {72 rotate 0 hpt L} repeat
  closepath stroke grestore Pnt } def
/PentF { stroke [] 0 setdash gsave
  translate 0 hpt M 4 {72 rotate 0 hpt L} repeat
  closepath fill grestore } def
/Circle { stroke [] 0 setdash 2 copy
  hpt 0 360 arc stroke Pnt } def
/CircleF { stroke [] 0 setdash hpt 0 360 arc fill } def
/C0 { BL [] 0 setdash 2 copy moveto vpt 90 450  arc } bind def
/C1 { BL [] 0 setdash 2 copy        moveto
       2 copy  vpt 0 90 arc closepath fill
               vpt 0 360 arc closepath } bind def
/C2 { BL [] 0 setdash 2 copy moveto
       2 copy  vpt 90 180 arc closepath fill
               vpt 0 360 arc closepath } bind def
/C3 { BL [] 0 setdash 2 copy moveto
       2 copy  vpt 0 180 arc closepath fill
               vpt 0 360 arc closepath } bind def
/C4 { BL [] 0 setdash 2 copy moveto
       2 copy  vpt 180 270 arc closepath fill
               vpt 0 360 arc closepath } bind def
/C5 { BL [] 0 setdash 2 copy moveto
       2 copy  vpt 0 90 arc
       2 copy moveto
       2 copy  vpt 180 270 arc closepath fill
               vpt 0 360 arc } bind def
/C6 { BL [] 0 setdash 2 copy moveto
      2 copy  vpt 90 270 arc closepath fill
              vpt 0 360 arc closepath } bind def
/C7 { BL [] 0 setdash 2 copy moveto
      2 copy  vpt 0 270 arc closepath fill
              vpt 0 360 arc closepath } bind def
/C8 { BL [] 0 setdash 2 copy moveto
      2 copy vpt 270 360 arc closepath fill
              vpt 0 360 arc closepath } bind def
/C9 { BL [] 0 setdash 2 copy moveto
      2 copy  vpt 270 450 arc closepath fill
              vpt 0 360 arc closepath } bind def
/C10 { BL [] 0 setdash 2 copy 2 copy moveto vpt 270 360 arc closepath fill
       2 copy moveto
       2 copy vpt 90 180 arc closepath fill
               vpt 0 360 arc closepath } bind def
/C11 { BL [] 0 setdash 2 copy moveto
       2 copy  vpt 0 180 arc closepath fill
       2 copy moveto
       2 copy  vpt 270 360 arc closepath fill
               vpt 0 360 arc closepath } bind def
/C12 { BL [] 0 setdash 2 copy moveto
       2 copy  vpt 180 360 arc closepath fill
               vpt 0 360 arc closepath } bind def
/C13 { BL [] 0 setdash  2 copy moveto
       2 copy  vpt 0 90 arc closepath fill
       2 copy moveto
       2 copy  vpt 180 360 arc closepath fill
               vpt 0 360 arc closepath } bind def
/C14 { BL [] 0 setdash 2 copy moveto
       2 copy  vpt 90 360 arc closepath fill
               vpt 0 360 arc } bind def
/C15 { BL [] 0 setdash 2 copy vpt 0 360 arc closepath fill
               vpt 0 360 arc closepath } bind def
/Rec   { newpath 4 2 roll moveto 1 index 0 rlineto 0 exch rlineto
       neg 0 rlineto closepath } bind def
/Square { dup Rec } bind def
/Bsquare { vpt sub exch vpt sub exch vpt2 Square } bind def
/S0 { BL [] 0 setdash 2 copy moveto 0 vpt rlineto BL Bsquare } bind def
/S1 { BL [] 0 setdash 2 copy vpt Square fill Bsquare } bind def
/S2 { BL [] 0 setdash 2 copy exch vpt sub exch vpt Square fill Bsquare } bind def
/S3 { BL [] 0 setdash 2 copy exch vpt sub exch vpt2 vpt Rec fill Bsquare } bind def
/S4 { BL [] 0 setdash 2 copy exch vpt sub exch vpt sub vpt Square fill Bsquare } bind def
/S5 { BL [] 0 setdash 2 copy 2 copy vpt Square fill
       exch vpt sub exch vpt sub vpt Square fill Bsquare } bind def
/S6 { BL [] 0 setdash 2 copy exch vpt sub exch vpt sub vpt vpt2 Rec fill Bsquare } bind def
/S7 { BL [] 0 setdash 2 copy exch vpt sub exch vpt sub vpt vpt2 Rec fill
       2 copy vpt Square fill
       Bsquare } bind def
/S8 { BL [] 0 setdash 2 copy vpt sub vpt Square fill Bsquare } bind def
/S9 { BL [] 0 setdash 2 copy vpt sub vpt vpt2 Rec fill Bsquare } bind def
/S10 { BL [] 0 setdash 2 copy vpt sub vpt Square fill 2 copy exch vpt sub exch vpt Square fill
       Bsquare } bind def
/S11 { BL [] 0 setdash 2 copy vpt sub vpt Square fill 2 copy exch vpt sub exch vpt2 vpt Rec fill
       Bsquare } bind def
/S12 { BL [] 0 setdash 2 copy exch vpt sub exch vpt sub vpt2 vpt Rec fill Bsquare } bind def
/S13 { BL [] 0 setdash 2 copy exch vpt sub exch vpt sub vpt2 vpt Rec fill
       2 copy vpt Square fill Bsquare } bind def
/S14 { BL [] 0 setdash 2 copy exch vpt sub exch vpt sub vpt2 vpt Rec fill
       2 copy exch vpt sub exch vpt Square fill Bsquare } bind def
/S15 { BL [] 0 setdash 2 copy Bsquare fill Bsquare } bind def
/D0 { gsave translate 45 rotate 0 0 S0 stroke grestore } bind def
/D1 { gsave translate 45 rotate 0 0 S1 stroke grestore } bind def
/D2 { gsave translate 45 rotate 0 0 S2 stroke grestore } bind def
/D3 { gsave translate 45 rotate 0 0 S3 stroke grestore } bind def
/D4 { gsave translate 45 rotate 0 0 S4 stroke grestore } bind def
/D5 { gsave translate 45 rotate 0 0 S5 stroke grestore } bind def
/D6 { gsave translate 45 rotate 0 0 S6 stroke grestore } bind def
/D7 { gsave translate 45 rotate 0 0 S7 stroke grestore } bind def
/D8 { gsave translate 45 rotate 0 0 S8 stroke grestore } bind def
/D9 { gsave translate 45 rotate 0 0 S9 stroke grestore } bind def
/D10 { gsave translate 45 rotate 0 0 S10 stroke grestore } bind def
/D11 { gsave translate 45 rotate 0 0 S11 stroke grestore } bind def
/D12 { gsave translate 45 rotate 0 0 S12 stroke grestore } bind def
/D13 { gsave translate 45 rotate 0 0 S13 stroke grestore } bind def
/D14 { gsave translate 45 rotate 0 0 S14 stroke grestore } bind def
/D15 { gsave translate 45 rotate 0 0 S15 stroke grestore } bind def
/DiaE { stroke [] 0 setdash vpt add M
  hpt neg vpt neg V hpt vpt neg V
  hpt vpt V hpt neg vpt V closepath stroke } def
/BoxE { stroke [] 0 setdash exch hpt sub exch vpt add M
  0 vpt2 neg V hpt2 0 V 0 vpt2 V
  hpt2 neg 0 V closepath stroke } def
/TriUE { stroke [] 0 setdash vpt 1.12 mul add M
  hpt neg vpt -1.62 mul V
  hpt 2 mul 0 V
  hpt neg vpt 1.62 mul V closepath stroke } def
/TriDE { stroke [] 0 setdash vpt 1.12 mul sub M
  hpt neg vpt 1.62 mul V
  hpt 2 mul 0 V
  hpt neg vpt -1.62 mul V closepath stroke } def
/PentE { stroke [] 0 setdash gsave
  translate 0 hpt M 4 {72 rotate 0 hpt L} repeat
  closepath stroke grestore } def
/CircE { stroke [] 0 setdash 
  hpt 0 360 arc stroke } def
/Opaque { gsave closepath 1 setgray fill grestore 0 setgray closepath } def
/DiaW { stroke [] 0 setdash vpt add M
  hpt neg vpt neg V hpt vpt neg V
  hpt vpt V hpt neg vpt V Opaque stroke } def
/BoxW { stroke [] 0 setdash exch hpt sub exch vpt add M
  0 vpt2 neg V hpt2 0 V 0 vpt2 V
  hpt2 neg 0 V Opaque stroke } def
/TriUW { stroke [] 0 setdash vpt 1.12 mul add M
  hpt neg vpt -1.62 mul V
  hpt 2 mul 0 V
  hpt neg vpt 1.62 mul V Opaque stroke } def
/TriDW { stroke [] 0 setdash vpt 1.12 mul sub M
  hpt neg vpt 1.62 mul V
  hpt 2 mul 0 V
  hpt neg vpt -1.62 mul V Opaque stroke } def
/PentW { stroke [] 0 setdash gsave
  translate 0 hpt M 4 {72 rotate 0 hpt L} repeat
  Opaque stroke grestore } def
/CircW { stroke [] 0 setdash 
  hpt 0 360 arc Opaque stroke } def
/BoxFill { gsave Rec 1 setgray fill grestore } def
/Symbol-Oblique /Symbol findfont [1 0 .167 1 0 0] makefont
dup length dict begin {1 index /FID eq {pop pop} {def} ifelse} forall
currentdict end definefont pop
end
}}%
\begin{picture}(2519,2160)(0,0)%
{\GNUPLOTspecial{"
gnudict begin
gsave
0 0 translate
0.100 0.100 scale
0 setgray
newpath
1.000 UL
LTb
500 1884 M
63 0 V
1807 0 R
-63 0 V
500 1532 M
63 0 V
1807 0 R
-63 0 V
500 1180 M
63 0 V
1807 0 R
-63 0 V
500 827 M
63 0 V
1807 0 R
-63 0 V
500 475 M
63 0 V
1807 0 R
-63 0 V
537 300 M
0 63 V
0 1697 R
0 -63 V
612 300 M
0 31 V
0 1729 R
0 -31 V
687 300 M
0 63 V
0 1697 R
0 -63 V
762 300 M
0 31 V
0 1729 R
0 -31 V
837 300 M
0 63 V
0 1697 R
0 -63 V
911 300 M
0 31 V
0 1729 R
0 -31 V
986 300 M
0 63 V
0 1697 R
0 -63 V
1061 300 M
0 31 V
0 1729 R
0 -31 V
1136 300 M
0 63 V
0 1697 R
0 -63 V
1211 300 M
0 31 V
0 1729 R
0 -31 V
1285 300 M
0 63 V
0 1697 R
0 -63 V
1360 300 M
0 31 V
0 1729 R
0 -31 V
1435 300 M
0 63 V
0 1697 R
0 -63 V
1510 300 M
0 31 V
0 1729 R
0 -31 V
1585 300 M
0 63 V
0 1697 R
0 -63 V
1659 300 M
0 31 V
0 1729 R
0 -31 V
1734 300 M
0 63 V
0 1697 R
0 -63 V
1809 300 M
0 31 V
0 1729 R
0 -31 V
1884 300 M
0 63 V
0 1697 R
0 -63 V
1959 300 M
0 31 V
0 1729 R
0 -31 V
2033 300 M
0 63 V
0 1697 R
0 -63 V
2108 300 M
0 31 V
0 1729 R
0 -31 V
2183 300 M
0 63 V
0 1697 R
0 -63 V
2258 300 M
0 31 V
0 1729 R
0 -31 V
2333 300 M
0 63 V
0 1697 R
0 -63 V
1.000 UL
LTb
500 300 M
1870 0 V
0 1760 V
-1870 0 V
500 300 L
0.500 UP
1.000 UL
LT3
537 726 Box
612 1206 Box
687 1694 Box
2154 1256 Box
0.500 UP
1.000 UL
LT4
537 592 BoxF
612 764 BoxF
687 968 BoxF
762 1213 BoxF
837 1525 BoxF
911 1784 BoxF
2154 1137 BoxF
0.500 UP
1.000 UL
LT5
537 514 Circle
612 591 Circle
687 680 Circle
762 786 Circle
837 908 Circle
911 1041 Circle
986 1179 Circle
1061 1320 Circle
1136 1461 Circle
1211 1605 Circle
1285 1771 Circle
2154 1018 Circle
0.500 UP
1.000 UL
LT6
537 461 CircleF
612 508 CircleF
687 559 CircleF
762 619 CircleF
837 686 CircleF
911 763 CircleF
986 849 CircleF
1061 940 CircleF
1136 1037 CircleF
1211 1142 CircleF
1285 1247 CircleF
1360 1343 CircleF
1435 1442 CircleF
1510 1545 CircleF
1585 1666 CircleF
1659 1786 CircleF
1173 1043 CircleF
2154 899 CircleF
0.500 UP
1.000 UL
LT7
537 420 TriU
612 452 TriU
687 487 TriU
762 524 TriU
837 566 TriU
911 613 TriU
986 664 TriU
1061 720 TriU
1136 784 TriU
1211 851 TriU
1285 921 TriU
1360 995 TriU
1435 1071 TriU
1510 1148 TriU
1585 1226 TriU
1659 1304 TriU
1734 1382 TriU
1809 1461 TriU
1884 1542 TriU
1959 1621 TriU
2033 1707 TriU
799 532 TriU
874 603 TriU
1098 729 TriU
1472 1061 TriU
2154 780 TriU
1.000 UL
LT1
2039 661 M
231 0 V
538 727 M
19 123 V
19 122 V
18 122 V
19 122 V
19 122 V
19 122 V
19 123 V
19 122 V
1.000 UL
LT1
613 735 M
19 66 V
19 65 V
19 66 V
19 65 V
19 66 V
19 66 V
19 65 V
18 66 V
19 65 V
19 66 V
19 66 V
19 65 V
19 66 V
19 65 V
19 66 V
19 66 V
1.000 UL
LT1
727 702 M
19 35 V
18 35 V
19 36 V
19 35 V
19 36 V
19 35 V
19 36 V
19 35 V
19 36 V
19 35 V
18 35 V
19 36 V
19 35 V
19 36 V
19 35 V
19 36 V
19 35 V
19 36 V
19 35 V
18 35 V
19 36 V
19 35 V
19 36 V
19 35 V
19 36 V
19 35 V
19 35 V
19 36 V
18 35 V
19 36 V
1.000 UL
LT1
897 721 M
19 26 V
18 26 V
19 25 V
19 26 V
19 26 V
19 26 V
19 25 V
19 26 V
19 26 V
19 26 V
18 25 V
19 26 V
19 26 V
19 26 V
19 25 V
19 26 V
19 26 V
19 25 V
19 26 V
18 26 V
19 26 V
19 25 V
19 26 V
19 26 V
19 26 V
19 25 V
19 26 V
19 26 V
18 26 V
19 25 V
19 26 V
19 26 V
19 25 V
19 26 V
19 26 V
19 26 V
19 25 V
18 26 V
19 26 V
19 26 V
19 25 V
1.000 UL
LT1
1123 756 M
19 19 V
19 20 V
19 19 V
19 20 V
19 19 V
19 20 V
19 19 V
18 20 V
19 19 V
19 20 V
19 19 V
19 20 V
19 19 V
19 20 V
19 19 V
19 20 V
18 19 V
19 20 V
19 19 V
19 20 V
19 19 V
19 20 V
19 19 V
19 20 V
19 19 V
18 20 V
19 19 V
19 20 V
19 19 V
19 20 V
19 19 V
19 20 V
19 19 V
19 20 V
18 19 V
19 20 V
19 19 V
19 20 V
19 19 V
19 20 V
19 19 V
19 20 V
19 19 V
18 20 V
19 19 V
19 20 V
19 19 V
19 20 V
1.000 UL
LT2
2039 542 M
231 0 V
538 617 M
19 0 V
19 0 V
18 0 V
19 0 V
19 0 V
19 0 V
19 0 V
19 0 V
19 0 V
19 0 V
19 0 V
18 0 V
19 0 V
19 0 V
19 0 V
19 0 V
19 0 V
19 0 V
19 0 V
19 0 V
18 0 V
19 0 V
19 0 V
19 0 V
19 0 V
19 0 V
19 0 V
19 0 V
19 0 V
18 0 V
19 0 V
1.000 UL
LT3
2039 423 M
231 0 V
689 1884 M
19 0 V
19 0 V
19 0 V
18 0 V
19 0 V
19 0 V
19 0 V
19 0 V
19 0 V
19 0 V
19 0 V
19 0 V
18 0 V
19 0 V
19 0 V
19 0 V
19 0 V
19 0 V
19 0 V
19 0 V
19 0 V
18 0 V
19 0 V
19 0 V
19 0 V
19 0 V
19 0 V
19 0 V
19 0 V
19 0 V
18 0 V
19 0 V
19 0 V
19 0 V
19 0 V
19 0 V
19 0 V
19 0 V
19 0 V
18 0 V
19 0 V
19 0 V
19 0 V
19 0 V
19 0 V
19 0 V
19 0 V
19 0 V
18 0 V
19 0 V
19 0 V
19 0 V
19 0 V
19 0 V
19 0 V
stroke
grestore
end
showpage
}}%
\put(1989,423){\makebox(0,0)[r]{$-{\hbar^2/(mb_t^2)}$}}%
\put(1989,542){\makebox(0,0)[r]{$-{\hbar^2/(mb^2)}$}}%
\put(1989,661){\makebox(0,0)[r]{$\mathrm{Const}\cdot e^{-\zeta_Nn}$}}%
\put(1989,780){\makebox(0,0)[r]{N=7}}%
\put(1989,899){\makebox(0,0)[r]{N=6}}%
\put(1989,1018){\makebox(0,0)[r]{N=5}}%
\put(1989,1137){\makebox(0,0)[r]{N=4}}%
\put(1989,1256){\makebox(0,0)[r]{N=3}}%
\put(1435,50){\makebox(0,0){$n$}}%
\put(200,1180){%
\special{ps: gsave currentpoint currentpoint translate
270 rotate neg exch neg exch translate}%
\makebox(0,0)[b]{\shortstack{$E_n/\hbar\omega$}}%
\special{ps: currentpoint grestore moveto}%
}%
\put(2333,200){\makebox(0,0){ 24}}%
\put(2183,200){\makebox(0,0){ 22}}%
\put(2033,200){\makebox(0,0){ 20}}%
\put(1884,200){\makebox(0,0){ 18}}%
\put(1734,200){\makebox(0,0){ 16}}%
\put(1585,200){\makebox(0,0){ 14}}%
\put(1435,200){\makebox(0,0){ 12}}%
\put(1285,200){\makebox(0,0){ 10}}%
\put(1136,200){\makebox(0,0){ 8}}%
\put(986,200){\makebox(0,0){ 6}}%
\put(837,200){\makebox(0,0){ 4}}%
\put(687,200){\makebox(0,0){ 2}}%
\put(537,200){\makebox(0,0){ 0}}%
\put(450,475){\makebox(0,0)[r]{$-10^8$}}%
\put(450,827){\makebox(0,0)[r]{$-10^6$}}%
\put(450,1180){\makebox(0,0)[r]{$-10^4$}}%
\put(450,1532){\makebox(0,0)[r]{$-10^2$}}%
\put(450,1884){\makebox(0,0)[r]{$-10^0$}}%
\end{picture}%
\endgroup
 

%% file: fig-z.tex
\begingroup%
  \makeatletter%
  \newcommand{\GNUPLOTspecial}{%
    \@sanitize\catcode`\%=14\relax\special}%
  \setlength{\unitlength}{0.1bp}%
{\GNUPLOTspecial{!
/gnudict 256 dict def
gnudict begin
/Color false def
/Solid false def
/gnulinewidth 5.000 def
/userlinewidth gnulinewidth def
/vshift -33 def
/dl {10 mul} def
/hpt_ 31.5 def
/vpt_ 31.5 def
/hpt hpt_ def
/vpt vpt_ def
/M {moveto} bind def
/L {lineto} bind def
/R {rmoveto} bind def
/V {rlineto} bind def
/vpt2 vpt 2 mul def
/hpt2 hpt 2 mul def
/Lshow { currentpoint stroke M
  0 vshift R show } def
/Rshow { currentpoint stroke M
  dup stringwidth pop neg vshift R show } def
/Cshow { currentpoint stroke M
  dup stringwidth pop -2 div vshift R show } def
/UP { dup vpt_ mul /vpt exch def hpt_ mul /hpt exch def
  /hpt2 hpt 2 mul def /vpt2 vpt 2 mul def } def
/DL { Color {setrgbcolor Solid {pop []} if 0 setdash }
 {pop pop pop Solid {pop []} if 0 setdash} ifelse } def
/BL { stroke userlinewidth 2 mul setlinewidth } def
/AL { stroke userlinewidth 2 div setlinewidth } def
/UL { dup gnulinewidth mul /userlinewidth exch def
      dup 1 lt {pop 1} if 10 mul /udl exch def } def
/PL { stroke userlinewidth setlinewidth } def
/LTb { BL [] 0 0 0 DL } def
/LTa { AL [1 udl mul 2 udl mul] 0 setdash 0 0 0 setrgbcolor } def
/LT0 { PL [] 1 0 0 DL } def
/LT1 { PL [4 dl 2 dl] 0 1 0 DL } def
/LT2 { PL [2 dl 3 dl] 0 0 1 DL } def
/LT3 { PL [1 dl 1.5 dl] 1 0 1 DL } def
/LT4 { PL [5 dl 2 dl 1 dl 2 dl] 0 1 1 DL } def
/LT5 { PL [4 dl 3 dl 1 dl 3 dl] 1 1 0 DL } def
/LT6 { PL [2 dl 2 dl 2 dl 4 dl] 0 0 0 DL } def
/LT7 { PL [2 dl 2 dl 2 dl 2 dl 2 dl 4 dl] 1 0.3 0 DL } def
/LT8 { PL [2 dl 2 dl 2 dl 2 dl 2 dl 2 dl 2 dl 4 dl] 0.5 0.5 0.5 DL } def
/Pnt { stroke [] 0 setdash
   gsave 1 setlinecap M 0 0 V stroke grestore } def
/Dia { stroke [] 0 setdash 2 copy vpt add M
  hpt neg vpt neg V hpt vpt neg V
  hpt vpt V hpt neg vpt V closepath stroke
  Pnt } def
/Pls { stroke [] 0 setdash vpt sub M 0 vpt2 V
  currentpoint stroke M
  hpt neg vpt neg R hpt2 0 V stroke
  } def
/Box { stroke [] 0 setdash 2 copy exch hpt sub exch vpt add M
  0 vpt2 neg V hpt2 0 V 0 vpt2 V
  hpt2 neg 0 V closepath stroke
  Pnt } def
/Crs { stroke [] 0 setdash exch hpt sub exch vpt add M
  hpt2 vpt2 neg V currentpoint stroke M
  hpt2 neg 0 R hpt2 vpt2 V stroke } def
/TriU { stroke [] 0 setdash 2 copy vpt 1.12 mul add M
  hpt neg vpt -1.62 mul V
  hpt 2 mul 0 V
  hpt neg vpt 1.62 mul V closepath stroke
  Pnt  } def
/Star { 2 copy Pls Crs } def
/BoxF { stroke [] 0 setdash exch hpt sub exch vpt add M
  0 vpt2 neg V  hpt2 0 V  0 vpt2 V
  hpt2 neg 0 V  closepath fill } def
/TriUF { stroke [] 0 setdash vpt 1.12 mul add M
  hpt neg vpt -1.62 mul V
  hpt 2 mul 0 V
  hpt neg vpt 1.62 mul V closepath fill } def
/TriD { stroke [] 0 setdash 2 copy vpt 1.12 mul sub M
  hpt neg vpt 1.62 mul V
  hpt 2 mul 0 V
  hpt neg vpt -1.62 mul V closepath stroke
  Pnt  } def
/TriDF { stroke [] 0 setdash vpt 1.12 mul sub M
  hpt neg vpt 1.62 mul V
  hpt 2 mul 0 V
  hpt neg vpt -1.62 mul V closepath fill} def
/DiaF { stroke [] 0 setdash vpt add M
  hpt neg vpt neg V hpt vpt neg V
  hpt vpt V hpt neg vpt V closepath fill } def
/Pent { stroke [] 0 setdash 2 copy gsave
  translate 0 hpt M 4 {72 rotate 0 hpt L} repeat
  closepath stroke grestore Pnt } def
/PentF { stroke [] 0 setdash gsave
  translate 0 hpt M 4 {72 rotate 0 hpt L} repeat
  closepath fill grestore } def
/Circle { stroke [] 0 setdash 2 copy
  hpt 0 360 arc stroke Pnt } def
/CircleF { stroke [] 0 setdash hpt 0 360 arc fill } def
/C0 { BL [] 0 setdash 2 copy moveto vpt 90 450  arc } bind def
/C1 { BL [] 0 setdash 2 copy        moveto
       2 copy  vpt 0 90 arc closepath fill
               vpt 0 360 arc closepath } bind def
/C2 { BL [] 0 setdash 2 copy moveto
       2 copy  vpt 90 180 arc closepath fill
               vpt 0 360 arc closepath } bind def
/C3 { BL [] 0 setdash 2 copy moveto
       2 copy  vpt 0 180 arc closepath fill
               vpt 0 360 arc closepath } bind def
/C4 { BL [] 0 setdash 2 copy moveto
       2 copy  vpt 180 270 arc closepath fill
               vpt 0 360 arc closepath } bind def
/C5 { BL [] 0 setdash 2 copy moveto
       2 copy  vpt 0 90 arc
       2 copy moveto
       2 copy  vpt 180 270 arc closepath fill
               vpt 0 360 arc } bind def
/C6 { BL [] 0 setdash 2 copy moveto
      2 copy  vpt 90 270 arc closepath fill
              vpt 0 360 arc closepath } bind def
/C7 { BL [] 0 setdash 2 copy moveto
      2 copy  vpt 0 270 arc closepath fill
              vpt 0 360 arc closepath } bind def
/C8 { BL [] 0 setdash 2 copy moveto
      2 copy vpt 270 360 arc closepath fill
              vpt 0 360 arc closepath } bind def
/C9 { BL [] 0 setdash 2 copy moveto
      2 copy  vpt 270 450 arc closepath fill
              vpt 0 360 arc closepath } bind def
/C10 { BL [] 0 setdash 2 copy 2 copy moveto vpt 270 360 arc closepath fill
       2 copy moveto
       2 copy vpt 90 180 arc closepath fill
               vpt 0 360 arc closepath } bind def
/C11 { BL [] 0 setdash 2 copy moveto
       2 copy  vpt 0 180 arc closepath fill
       2 copy moveto
       2 copy  vpt 270 360 arc closepath fill
               vpt 0 360 arc closepath } bind def
/C12 { BL [] 0 setdash 2 copy moveto
       2 copy  vpt 180 360 arc closepath fill
               vpt 0 360 arc closepath } bind def
/C13 { BL [] 0 setdash  2 copy moveto
       2 copy  vpt 0 90 arc closepath fill
       2 copy moveto
       2 copy  vpt 180 360 arc closepath fill
               vpt 0 360 arc closepath } bind def
/C14 { BL [] 0 setdash 2 copy moveto
       2 copy  vpt 90 360 arc closepath fill
               vpt 0 360 arc } bind def
/C15 { BL [] 0 setdash 2 copy vpt 0 360 arc closepath fill
               vpt 0 360 arc closepath } bind def
/Rec   { newpath 4 2 roll moveto 1 index 0 rlineto 0 exch rlineto
       neg 0 rlineto closepath } bind def
/Square { dup Rec } bind def
/Bsquare { vpt sub exch vpt sub exch vpt2 Square } bind def
/S0 { BL [] 0 setdash 2 copy moveto 0 vpt rlineto BL Bsquare } bind def
/S1 { BL [] 0 setdash 2 copy vpt Square fill Bsquare } bind def
/S2 { BL [] 0 setdash 2 copy exch vpt sub exch vpt Square fill Bsquare } bind def
/S3 { BL [] 0 setdash 2 copy exch vpt sub exch vpt2 vpt Rec fill Bsquare } bind def
/S4 { BL [] 0 setdash 2 copy exch vpt sub exch vpt sub vpt Square fill Bsquare } bind def
/S5 { BL [] 0 setdash 2 copy 2 copy vpt Square fill
       exch vpt sub exch vpt sub vpt Square fill Bsquare } bind def
/S6 { BL [] 0 setdash 2 copy exch vpt sub exch vpt sub vpt vpt2 Rec fill Bsquare } bind def
/S7 { BL [] 0 setdash 2 copy exch vpt sub exch vpt sub vpt vpt2 Rec fill
       2 copy vpt Square fill
       Bsquare } bind def
/S8 { BL [] 0 setdash 2 copy vpt sub vpt Square fill Bsquare } bind def
/S9 { BL [] 0 setdash 2 copy vpt sub vpt vpt2 Rec fill Bsquare } bind def
/S10 { BL [] 0 setdash 2 copy vpt sub vpt Square fill 2 copy exch vpt sub exch vpt Square fill
       Bsquare } bind def
/S11 { BL [] 0 setdash 2 copy vpt sub vpt Square fill 2 copy exch vpt sub exch vpt2 vpt Rec fill
       Bsquare } bind def
/S12 { BL [] 0 setdash 2 copy exch vpt sub exch vpt sub vpt2 vpt Rec fill Bsquare } bind def
/S13 { BL [] 0 setdash 2 copy exch vpt sub exch vpt sub vpt2 vpt Rec fill
       2 copy vpt Square fill Bsquare } bind def
/S14 { BL [] 0 setdash 2 copy exch vpt sub exch vpt sub vpt2 vpt Rec fill
       2 copy exch vpt sub exch vpt Square fill Bsquare } bind def
/S15 { BL [] 0 setdash 2 copy Bsquare fill Bsquare } bind def
/D0 { gsave translate 45 rotate 0 0 S0 stroke grestore } bind def
/D1 { gsave translate 45 rotate 0 0 S1 stroke grestore } bind def
/D2 { gsave translate 45 rotate 0 0 S2 stroke grestore } bind def
/D3 { gsave translate 45 rotate 0 0 S3 stroke grestore } bind def
/D4 { gsave translate 45 rotate 0 0 S4 stroke grestore } bind def
/D5 { gsave translate 45 rotate 0 0 S5 stroke grestore } bind def
/D6 { gsave translate 45 rotate 0 0 S6 stroke grestore } bind def
/D7 { gsave translate 45 rotate 0 0 S7 stroke grestore } bind def
/D8 { gsave translate 45 rotate 0 0 S8 stroke grestore } bind def
/D9 { gsave translate 45 rotate 0 0 S9 stroke grestore } bind def
/D10 { gsave translate 45 rotate 0 0 S10 stroke grestore } bind def
/D11 { gsave translate 45 rotate 0 0 S11 stroke grestore } bind def
/D12 { gsave translate 45 rotate 0 0 S12 stroke grestore } bind def
/D13 { gsave translate 45 rotate 0 0 S13 stroke grestore } bind def
/D14 { gsave translate 45 rotate 0 0 S14 stroke grestore } bind def
/D15 { gsave translate 45 rotate 0 0 S15 stroke grestore } bind def
/DiaE { stroke [] 0 setdash vpt add M
  hpt neg vpt neg V hpt vpt neg V
  hpt vpt V hpt neg vpt V closepath stroke } def
/BoxE { stroke [] 0 setdash exch hpt sub exch vpt add M
  0 vpt2 neg V hpt2 0 V 0 vpt2 V
  hpt2 neg 0 V closepath stroke } def
/TriUE { stroke [] 0 setdash vpt 1.12 mul add M
  hpt neg vpt -1.62 mul V
  hpt 2 mul 0 V
  hpt neg vpt 1.62 mul V closepath stroke } def
/TriDE { stroke [] 0 setdash vpt 1.12 mul sub M
  hpt neg vpt 1.62 mul V
  hpt 2 mul 0 V
  hpt neg vpt -1.62 mul V closepath stroke } def
/PentE { stroke [] 0 setdash gsave
  translate 0 hpt M 4 {72 rotate 0 hpt L} repeat
  closepath stroke grestore } def
/CircE { stroke [] 0 setdash 
  hpt 0 360 arc stroke } def
/Opaque { gsave closepath 1 setgray fill grestore 0 setgray closepath } def
/DiaW { stroke [] 0 setdash vpt add M
  hpt neg vpt neg V hpt vpt neg V
  hpt vpt V hpt neg vpt V Opaque stroke } def
/BoxW { stroke [] 0 setdash exch hpt sub exch vpt add M
  0 vpt2 neg V hpt2 0 V 0 vpt2 V
  hpt2 neg 0 V Opaque stroke } def
/TriUW { stroke [] 0 setdash vpt 1.12 mul add M
  hpt neg vpt -1.62 mul V
  hpt 2 mul 0 V
  hpt neg vpt 1.62 mul V Opaque stroke } def
/TriDW { stroke [] 0 setdash vpt 1.12 mul sub M
  hpt neg vpt 1.62 mul V
  hpt 2 mul 0 V
  hpt neg vpt -1.62 mul V Opaque stroke } def
/PentW { stroke [] 0 setdash gsave
  translate 0 hpt M 4 {72 rotate 0 hpt L} repeat
  Opaque stroke grestore } def
/CircW { stroke [] 0 setdash 
  hpt 0 360 arc Opaque stroke } def
/BoxFill { gsave Rec 1 setgray fill grestore } def
/Symbol-Oblique /Symbol findfont [1 0 .167 1 0 0] makefont
dup length dict begin {1 index /FID eq {pop pop} {def} ifelse} forall
currentdict end definefont pop
end
}}%
\begin{picture}(2519,2160)(0,0)%
{\GNUPLOTspecial{"
gnudict begin
gsave
0 0 translate
0.100 0.100 scale
0 setgray
newpath
1.000 UL
LTb
350 300 M
63 0 V
1957 0 R
-63 0 V
350 426 M
31 0 V
1989 0 R
-31 0 V
350 551 M
63 0 V
1957 0 R
-63 0 V
350 677 M
31 0 V
1989 0 R
-31 0 V
350 803 M
63 0 V
1957 0 R
-63 0 V
350 929 M
31 0 V
1989 0 R
-31 0 V
350 1054 M
63 0 V
1957 0 R
-63 0 V
350 1180 M
31 0 V
1989 0 R
-31 0 V
350 1306 M
63 0 V
1957 0 R
-63 0 V
350 1431 M
31 0 V
1989 0 R
-31 0 V
350 1557 M
63 0 V
1957 0 R
-63 0 V
350 1683 M
31 0 V
1989 0 R
-31 0 V
350 1809 M
63 0 V
1957 0 R
-63 0 V
350 1934 M
31 0 V
1989 0 R
-31 0 V
350 2060 M
63 0 V
1957 0 R
-63 0 V
552 300 M
0 63 V
0 1697 R
0 -63 V
956 300 M
0 63 V
0 1697 R
0 -63 V
1360 300 M
0 63 V
0 1697 R
0 -63 V
1764 300 M
0 63 V
0 1697 R
0 -63 V
2168 300 M
0 63 V
0 1697 R
0 -63 V
1.000 UL
LTb
350 300 M
2020 0 V
0 1760 V
-2020 0 V
350 300 L
0.500 UP
1.000 UL
LT0
552 1883 M
0 16 V
-31 -16 R
62 0 V
-62 16 R
62 0 V
956 1119 M
0 70 V
-31 -70 R
62 0 V
-62 70 R
62 0 V
1360 755 M
0 13 V
-31 -13 R
62 0 V
-62 13 R
62 0 V
1764 631 M
0 8 V
-31 -8 R
62 0 V
-62 8 R
62 0 V
373 -87 R
0 4 V
-31 -4 R
62 0 V
-62 4 R
62 0 V
552 1891 Pls
956 1154 Pls
1360 761 Pls
1764 635 Pls
2168 554 Pls
0.500 UP
1.000 UL
LT0
2039 1938 M
231 0 V
552 1891 M
956 1154 L
1360 761 L
1764 635 L
404 -81 V
552 1891 BoxF
956 1154 BoxF
1360 761 BoxF
1764 635 BoxF
2168 554 BoxF
2154 1938 BoxF
1.000 UL
LT2
2039 1819 M
231 0 V
962 826 M
21 -10 V
20 -10 V
20 -10 V
21 -9 V
20 -9 V
21 -9 V
20 -8 V
20 -9 V
21 -7 V
20 -8 V
21 -7 V
20 -8 V
20 -7 V
21 -6 V
20 -7 V
21 -6 V
20 -6 V
20 -6 V
21 -6 V
20 -6 V
21 -5 V
20 -6 V
20 -5 V
21 -5 V
20 -5 V
21 -5 V
20 -5 V
20 -4 V
21 -5 V
20 -4 V
21 -5 V
20 -4 V
20 -4 V
21 -4 V
20 -4 V
21 -4 V
20 -3 V
20 -4 V
21 -4 V
20 -3 V
21 -4 V
20 -3 V
20 -3 V
21 -3 V
20 -4 V
21 -3 V
20 -3 V
21 -3 V
20 -3 V
20 -2 V
21 -3 V
20 -3 V
21 -3 V
20 -2 V
20 -3 V
21 -2 V
20 -3 V
21 -2 V
20 -3 V
20 -2 V
21 -2 V
20 -3 V
21 -2 V
20 -2 V
20 -2 V
21 -2 V
20 -3 V
21 -2 V
20 -2 V
stroke
grestore
end
showpage
}}%
\put(1989,1819){\makebox(0,0)[r]{asymptotic form eq.~(\ref{eq:zN})}}%
\put(1989,1938){\makebox(0,0)[r]{$\zeta_N$ from Fig.~\ref{fig:e}}}%
\put(1360,50){\makebox(0,0){$N$}}%
\put(100,1180){%
\special{ps: gsave currentpoint currentpoint translate
270 rotate neg exch neg exch translate}%
\makebox(0,0)[b]{\shortstack{$\zeta_N$}}%
\special{ps: currentpoint grestore moveto}%
}%
\put(2168,200){\makebox(0,0){ 7}}%
\put(1764,200){\makebox(0,0){ 6}}%
\put(1360,200){\makebox(0,0){ 5}}%
\put(956,200){\makebox(0,0){ 4}}%
\put(552,200){\makebox(0,0){ 3}}%
\put(300,2060){\makebox(0,0)[r]{ 7}}%
\put(300,1809){\makebox(0,0)[r]{ 6}}%
\put(300,1557){\makebox(0,0)[r]{ 5}}%
\put(300,1306){\makebox(0,0)[r]{ 4}}%
\put(300,1054){\makebox(0,0)[r]{ 3}}%
\put(300,803){\makebox(0,0)[r]{ 2}}%
\put(300,551){\makebox(0,0)[r]{ 1}}%
\put(300,300){\makebox(0,0)[r]{ 0}}%
\end{picture}%
\endgroup
 

%% file: fig-r.tex
\begingroup%
  \makeatletter%
  \newcommand{\GNUPLOTspecial}{%
    \@sanitize\catcode`\%=14\relax\special}%
  \setlength{\unitlength}{0.1bp}%
{\GNUPLOTspecial{!
/gnudict 256 dict def
gnudict begin
/Color false def
/Solid false def
/gnulinewidth 5.000 def
/userlinewidth gnulinewidth def
/vshift -33 def
/dl {10 mul} def
/hpt_ 31.5 def
/vpt_ 31.5 def
/hpt hpt_ def
/vpt vpt_ def
/M {moveto} bind def
/L {lineto} bind def
/R {rmoveto} bind def
/V {rlineto} bind def
/vpt2 vpt 2 mul def
/hpt2 hpt 2 mul def
/Lshow { currentpoint stroke M
  0 vshift R show } def
/Rshow { currentpoint stroke M
  dup stringwidth pop neg vshift R show } def
/Cshow { currentpoint stroke M
  dup stringwidth pop -2 div vshift R show } def
/UP { dup vpt_ mul /vpt exch def hpt_ mul /hpt exch def
  /hpt2 hpt 2 mul def /vpt2 vpt 2 mul def } def
/DL { Color {setrgbcolor Solid {pop []} if 0 setdash }
 {pop pop pop Solid {pop []} if 0 setdash} ifelse } def
/BL { stroke userlinewidth 2 mul setlinewidth } def
/AL { stroke userlinewidth 2 div setlinewidth } def
/UL { dup gnulinewidth mul /userlinewidth exch def
      dup 1 lt {pop 1} if 10 mul /udl exch def } def
/PL { stroke userlinewidth setlinewidth } def
/LTb { BL [] 0 0 0 DL } def
/LTa { AL [1 udl mul 2 udl mul] 0 setdash 0 0 0 setrgbcolor } def
/LT0 { PL [] 1 0 0 DL } def
/LT1 { PL [4 dl 2 dl] 0 1 0 DL } def
/LT2 { PL [2 dl 3 dl] 0 0 1 DL } def
/LT3 { PL [1 dl 1.5 dl] 1 0 1 DL } def
/LT4 { PL [5 dl 2 dl 1 dl 2 dl] 0 1 1 DL } def
/LT5 { PL [4 dl 3 dl 1 dl 3 dl] 1 1 0 DL } def
/LT6 { PL [2 dl 2 dl 2 dl 4 dl] 0 0 0 DL } def
/LT7 { PL [2 dl 2 dl 2 dl 2 dl 2 dl 4 dl] 1 0.3 0 DL } def
/LT8 { PL [2 dl 2 dl 2 dl 2 dl 2 dl 2 dl 2 dl 4 dl] 0.5 0.5 0.5 DL } def
/Pnt { stroke [] 0 setdash
   gsave 1 setlinecap M 0 0 V stroke grestore } def
/Dia { stroke [] 0 setdash 2 copy vpt add M
  hpt neg vpt neg V hpt vpt neg V
  hpt vpt V hpt neg vpt V closepath stroke
  Pnt } def
/Pls { stroke [] 0 setdash vpt sub M 0 vpt2 V
  currentpoint stroke M
  hpt neg vpt neg R hpt2 0 V stroke
  } def
/Box { stroke [] 0 setdash 2 copy exch hpt sub exch vpt add M
  0 vpt2 neg V hpt2 0 V 0 vpt2 V
  hpt2 neg 0 V closepath stroke
  Pnt } def
/Crs { stroke [] 0 setdash exch hpt sub exch vpt add M
  hpt2 vpt2 neg V currentpoint stroke M
  hpt2 neg 0 R hpt2 vpt2 V stroke } def
/TriU { stroke [] 0 setdash 2 copy vpt 1.12 mul add M
  hpt neg vpt -1.62 mul V
  hpt 2 mul 0 V
  hpt neg vpt 1.62 mul V closepath stroke
  Pnt  } def
/Star { 2 copy Pls Crs } def
/BoxF { stroke [] 0 setdash exch hpt sub exch vpt add M
  0 vpt2 neg V  hpt2 0 V  0 vpt2 V
  hpt2 neg 0 V  closepath fill } def
/TriUF { stroke [] 0 setdash vpt 1.12 mul add M
  hpt neg vpt -1.62 mul V
  hpt 2 mul 0 V
  hpt neg vpt 1.62 mul V closepath fill } def
/TriD { stroke [] 0 setdash 2 copy vpt 1.12 mul sub M
  hpt neg vpt 1.62 mul V
  hpt 2 mul 0 V
  hpt neg vpt -1.62 mul V closepath stroke
  Pnt  } def
/TriDF { stroke [] 0 setdash vpt 1.12 mul sub M
  hpt neg vpt 1.62 mul V
  hpt 2 mul 0 V
  hpt neg vpt -1.62 mul V closepath fill} def
/DiaF { stroke [] 0 setdash vpt add M
  hpt neg vpt neg V hpt vpt neg V
  hpt vpt V hpt neg vpt V closepath fill } def
/Pent { stroke [] 0 setdash 2 copy gsave
  translate 0 hpt M 4 {72 rotate 0 hpt L} repeat
  closepath stroke grestore Pnt } def
/PentF { stroke [] 0 setdash gsave
  translate 0 hpt M 4 {72 rotate 0 hpt L} repeat
  closepath fill grestore } def
/Circle { stroke [] 0 setdash 2 copy
  hpt 0 360 arc stroke Pnt } def
/CircleF { stroke [] 0 setdash hpt 0 360 arc fill } def
/C0 { BL [] 0 setdash 2 copy moveto vpt 90 450  arc } bind def
/C1 { BL [] 0 setdash 2 copy        moveto
       2 copy  vpt 0 90 arc closepath fill
               vpt 0 360 arc closepath } bind def
/C2 { BL [] 0 setdash 2 copy moveto
       2 copy  vpt 90 180 arc closepath fill
               vpt 0 360 arc closepath } bind def
/C3 { BL [] 0 setdash 2 copy moveto
       2 copy  vpt 0 180 arc closepath fill
               vpt 0 360 arc closepath } bind def
/C4 { BL [] 0 setdash 2 copy moveto
       2 copy  vpt 180 270 arc closepath fill
               vpt 0 360 arc closepath } bind def
/C5 { BL [] 0 setdash 2 copy moveto
       2 copy  vpt 0 90 arc
       2 copy moveto
       2 copy  vpt 180 270 arc closepath fill
               vpt 0 360 arc } bind def
/C6 { BL [] 0 setdash 2 copy moveto
      2 copy  vpt 90 270 arc closepath fill
              vpt 0 360 arc closepath } bind def
/C7 { BL [] 0 setdash 2 copy moveto
      2 copy  vpt 0 270 arc closepath fill
              vpt 0 360 arc closepath } bind def
/C8 { BL [] 0 setdash 2 copy moveto
      2 copy vpt 270 360 arc closepath fill
              vpt 0 360 arc closepath } bind def
/C9 { BL [] 0 setdash 2 copy moveto
      2 copy  vpt 270 450 arc closepath fill
              vpt 0 360 arc closepath } bind def
/C10 { BL [] 0 setdash 2 copy 2 copy moveto vpt 270 360 arc closepath fill
       2 copy moveto
       2 copy vpt 90 180 arc closepath fill
               vpt 0 360 arc closepath } bind def
/C11 { BL [] 0 setdash 2 copy moveto
       2 copy  vpt 0 180 arc closepath fill
       2 copy moveto
       2 copy  vpt 270 360 arc closepath fill
               vpt 0 360 arc closepath } bind def
/C12 { BL [] 0 setdash 2 copy moveto
       2 copy  vpt 180 360 arc closepath fill
               vpt 0 360 arc closepath } bind def
/C13 { BL [] 0 setdash  2 copy moveto
       2 copy  vpt 0 90 arc closepath fill
       2 copy moveto
       2 copy  vpt 180 360 arc closepath fill
               vpt 0 360 arc closepath } bind def
/C14 { BL [] 0 setdash 2 copy moveto
       2 copy  vpt 90 360 arc closepath fill
               vpt 0 360 arc } bind def
/C15 { BL [] 0 setdash 2 copy vpt 0 360 arc closepath fill
               vpt 0 360 arc closepath } bind def
/Rec   { newpath 4 2 roll moveto 1 index 0 rlineto 0 exch rlineto
       neg 0 rlineto closepath } bind def
/Square { dup Rec } bind def
/Bsquare { vpt sub exch vpt sub exch vpt2 Square } bind def
/S0 { BL [] 0 setdash 2 copy moveto 0 vpt rlineto BL Bsquare } bind def
/S1 { BL [] 0 setdash 2 copy vpt Square fill Bsquare } bind def
/S2 { BL [] 0 setdash 2 copy exch vpt sub exch vpt Square fill Bsquare } bind def
/S3 { BL [] 0 setdash 2 copy exch vpt sub exch vpt2 vpt Rec fill Bsquare } bind def
/S4 { BL [] 0 setdash 2 copy exch vpt sub exch vpt sub vpt Square fill Bsquare } bind def
/S5 { BL [] 0 setdash 2 copy 2 copy vpt Square fill
       exch vpt sub exch vpt sub vpt Square fill Bsquare } bind def
/S6 { BL [] 0 setdash 2 copy exch vpt sub exch vpt sub vpt vpt2 Rec fill Bsquare } bind def
/S7 { BL [] 0 setdash 2 copy exch vpt sub exch vpt sub vpt vpt2 Rec fill
       2 copy vpt Square fill
       Bsquare } bind def
/S8 { BL [] 0 setdash 2 copy vpt sub vpt Square fill Bsquare } bind def
/S9 { BL [] 0 setdash 2 copy vpt sub vpt vpt2 Rec fill Bsquare } bind def
/S10 { BL [] 0 setdash 2 copy vpt sub vpt Square fill 2 copy exch vpt sub exch vpt Square fill
       Bsquare } bind def
/S11 { BL [] 0 setdash 2 copy vpt sub vpt Square fill 2 copy exch vpt sub exch vpt2 vpt Rec fill
       Bsquare } bind def
/S12 { BL [] 0 setdash 2 copy exch vpt sub exch vpt sub vpt2 vpt Rec fill Bsquare } bind def
/S13 { BL [] 0 setdash 2 copy exch vpt sub exch vpt sub vpt2 vpt Rec fill
       2 copy vpt Square fill Bsquare } bind def
/S14 { BL [] 0 setdash 2 copy exch vpt sub exch vpt sub vpt2 vpt Rec fill
       2 copy exch vpt sub exch vpt Square fill Bsquare } bind def
/S15 { BL [] 0 setdash 2 copy Bsquare fill Bsquare } bind def
/D0 { gsave translate 45 rotate 0 0 S0 stroke grestore } bind def
/D1 { gsave translate 45 rotate 0 0 S1 stroke grestore } bind def
/D2 { gsave translate 45 rotate 0 0 S2 stroke grestore } bind def
/D3 { gsave translate 45 rotate 0 0 S3 stroke grestore } bind def
/D4 { gsave translate 45 rotate 0 0 S4 stroke grestore } bind def
/D5 { gsave translate 45 rotate 0 0 S5 stroke grestore } bind def
/D6 { gsave translate 45 rotate 0 0 S6 stroke grestore } bind def
/D7 { gsave translate 45 rotate 0 0 S7 stroke grestore } bind def
/D8 { gsave translate 45 rotate 0 0 S8 stroke grestore } bind def
/D9 { gsave translate 45 rotate 0 0 S9 stroke grestore } bind def
/D10 { gsave translate 45 rotate 0 0 S10 stroke grestore } bind def
/D11 { gsave translate 45 rotate 0 0 S11 stroke grestore } bind def
/D12 { gsave translate 45 rotate 0 0 S12 stroke grestore } bind def
/D13 { gsave translate 45 rotate 0 0 S13 stroke grestore } bind def
/D14 { gsave translate 45 rotate 0 0 S14 stroke grestore } bind def
/D15 { gsave translate 45 rotate 0 0 S15 stroke grestore } bind def
/DiaE { stroke [] 0 setdash vpt add M
  hpt neg vpt neg V hpt vpt neg V
  hpt vpt V hpt neg vpt V closepath stroke } def
/BoxE { stroke [] 0 setdash exch hpt sub exch vpt add M
  0 vpt2 neg V hpt2 0 V 0 vpt2 V
  hpt2 neg 0 V closepath stroke } def
/TriUE { stroke [] 0 setdash vpt 1.12 mul add M
  hpt neg vpt -1.62 mul V
  hpt 2 mul 0 V
  hpt neg vpt 1.62 mul V closepath stroke } def
/TriDE { stroke [] 0 setdash vpt 1.12 mul sub M
  hpt neg vpt 1.62 mul V
  hpt 2 mul 0 V
  hpt neg vpt -1.62 mul V closepath stroke } def
/PentE { stroke [] 0 setdash gsave
  translate 0 hpt M 4 {72 rotate 0 hpt L} repeat
  closepath stroke grestore } def
/CircE { stroke [] 0 setdash 
  hpt 0 360 arc stroke } def
/Opaque { gsave closepath 1 setgray fill grestore 0 setgray closepath } def
/DiaW { stroke [] 0 setdash vpt add M
  hpt neg vpt neg V hpt vpt neg V
  hpt vpt V hpt neg vpt V Opaque stroke } def
/BoxW { stroke [] 0 setdash exch hpt sub exch vpt add M
  0 vpt2 neg V hpt2 0 V 0 vpt2 V
  hpt2 neg 0 V Opaque stroke } def
/TriUW { stroke [] 0 setdash vpt 1.12 mul add M
  hpt neg vpt -1.62 mul V
  hpt 2 mul 0 V
  hpt neg vpt 1.62 mul V Opaque stroke } def
/TriDW { stroke [] 0 setdash vpt 1.12 mul sub M
  hpt neg vpt 1.62 mul V
  hpt 2 mul 0 V
  hpt neg vpt -1.62 mul V Opaque stroke } def
/PentW { stroke [] 0 setdash gsave
  translate 0 hpt M 4 {72 rotate 0 hpt L} repeat
  Opaque stroke grestore } def
/CircW { stroke [] 0 setdash 
  hpt 0 360 arc Opaque stroke } def
/BoxFill { gsave Rec 1 setgray fill grestore } def
/Symbol-Oblique /Symbol findfont [1 0 .167 1 0 0] makefont
dup length dict begin {1 index /FID eq {pop pop} {def} ifelse} forall
currentdict end definefont pop
end
}}%
\begin{picture}(2519,2160)(0,0)%
{\GNUPLOTspecial{"
gnudict begin
gsave
0 0 translate
0.100 0.100 scale
0 setgray
newpath
1.000 UL
LTb
600 2060 M
63 0 V
1707 0 R
-63 0 V
600 1685 M
63 0 V
1707 0 R
-63 0 V
600 1311 M
63 0 V
1707 0 R
-63 0 V
600 936 M
63 0 V
1707 0 R
-63 0 V
600 562 M
63 0 V
1707 0 R
-63 0 V
662 300 M
0 63 V
0 1697 R
0 -63 V
730 300 M
0 31 V
0 1729 R
0 -31 V
799 300 M
0 63 V
0 1697 R
0 -63 V
868 300 M
0 31 V
0 1729 R
0 -31 V
936 300 M
0 63 V
0 1697 R
0 -63 V
1005 300 M
0 31 V
0 1729 R
0 -31 V
1073 300 M
0 63 V
0 1697 R
0 -63 V
1142 300 M
0 31 V
0 1729 R
0 -31 V
1211 300 M
0 63 V
0 1697 R
0 -63 V
1279 300 M
0 31 V
0 1729 R
0 -31 V
1348 300 M
0 63 V
0 1697 R
0 -63 V
1416 300 M
0 31 V
0 1729 R
0 -31 V
1485 300 M
0 63 V
0 1697 R
0 -63 V
1554 300 M
0 31 V
0 1729 R
0 -31 V
1622 300 M
0 63 V
0 1697 R
0 -63 V
1691 300 M
0 31 V
0 1729 R
0 -31 V
1759 300 M
0 63 V
0 1697 R
0 -63 V
1828 300 M
0 31 V
0 1729 R
0 -31 V
1897 300 M
0 63 V
0 1697 R
0 -63 V
1965 300 M
0 31 V
0 1729 R
0 -31 V
2034 300 M
0 63 V
0 1697 R
0 -63 V
2102 300 M
0 31 V
0 1729 R
0 -31 V
2171 300 M
0 63 V
0 1697 R
0 -63 V
2240 300 M
0 31 V
0 1729 R
0 -31 V
2308 300 M
0 63 V
0 1697 R
0 -63 V
1.000 UL
LTb
600 300 M
1770 0 V
0 1760 V
-1770 0 V
600 300 L
0.500 UP
1.000 UL
LT3
662 458 Box
730 933 Box
799 1440 Box
868 1718 Box
936 1799 Box
1005 1828 Box
1073 1832 Box
1142 1837 Box
1211 1841 Box
1279 1861 Box
1348 1840 Box
2154 1018 Box
0.500 UP
1.000 UL
LT4
662 424 BoxF
730 589 BoxF
799 788 BoxF
868 1028 BoxF
936 1363 BoxF
1005 1596 BoxF
1073 1728 BoxF
1142 1823 BoxF
1211 1846 BoxF
1279 1858 BoxF
1348 1864 BoxF
2154 899 BoxF
0.500 UP
1.000 UL
LT5
662 417 Circle
730 501 Circle
799 588 Circle
868 684 Circle
936 796 Circle
1005 923 Circle
1073 1061 Circle
1142 1205 Circle
1211 1353 Circle
1279 1503 Circle
1348 1649 Circle
1416 1734 Circle
1485 1831 Circle
1554 1846 Circle
1622 1868 Circle
1691 1873 Circle
1759 1886 Circle
1828 1887 Circle
1897 1890 Circle
1965 1898 Circle
2034 1901 Circle
2102 1903 Circle
2171 1899 Circle
2240 1910 Circle
2308 1909 Circle
2154 780 Circle
0.500 UP
1.000 UL
LT6
662 417 CircleF
730 474 CircleF
799 531 CircleF
868 588 CircleF
936 649 CircleF
1005 718 CircleF
1073 793 CircleF
1142 878 CircleF
1211 953 CircleF
1279 1071 CircleF
1348 1171 CircleF
1416 1274 CircleF
1485 1382 CircleF
1554 1491 CircleF
1622 1594 CircleF
1691 1706 CircleF
1828 1787 CircleF
1897 1833 CircleF
1965 1855 CircleF
2034 1862 CircleF
1245 837 CircleF
2154 661 CircleF
0.500 UP
1.000 UL
LT7
662 418 TriU
730 461 TriU
799 503 TriU
868 543 TriU
936 584 TriU
1005 627 TriU
1073 673 TriU
1142 716 TriU
1211 777 TriU
1279 837 TriU
1348 901 TriU
1416 970 TriU
1485 1041 TriU
1554 1120 TriU
1622 1199 TriU
1691 1278 TriU
1759 1361 TriU
1828 1443 TriU
1897 1525 TriU
1965 1604 TriU
2034 1687 TriU
902 494 TriU
970 551 TriU
1176 633 TriU
1519 768 TriU
2154 542 TriU
1.000 UL
LT1
2039 423 M
231 0 V
654 369 M
18 134 V
17 134 V
18 134 V
18 134 V
18 134 V
18 134 V
18 134 V
18 134 V
1.000 UL
LT1
725 499 M
18 72 V
18 72 V
18 72 V
18 72 V
18 72 V
17 72 V
18 72 V
18 72 V
18 72 V
18 72 V
18 72 V
18 72 V
18 72 V
17 72 V
18 72 V
18 72 V
1.000 UL
LT1
850 587 M
18 39 V
18 38 V
18 39 V
18 39 V
18 39 V
18 39 V
17 39 V
18 39 V
18 39 V
18 39 V
18 39 V
18 38 V
18 39 V
18 39 V
17 39 V
18 39 V
18 39 V
18 39 V
18 39 V
18 39 V
18 39 V
18 38 V
18 39 V
17 39 V
18 39 V
18 39 V
18 39 V
18 39 V
1.000 UL
LT1
993 609 M
18 28 V
18 29 V
18 28 V
18 28 V
18 28 V
18 29 V
17 28 V
18 28 V
18 28 V
18 29 V
18 28 V
18 28 V
18 28 V
18 29 V
18 28 V
17 28 V
18 28 V
18 29 V
18 28 V
18 28 V
18 28 V
18 29 V
18 28 V
17 28 V
18 28 V
18 29 V
18 28 V
18 28 V
18 28 V
18 29 V
18 28 V
17 28 V
18 28 V
18 29 V
18 28 V
18 28 V
18 28 V
18 29 V
18 28 V
1.000 UL
LT1
1208 692 M
18 21 V
18 22 V
18 21 V
17 22 V
18 21 V
18 21 V
18 22 V
18 21 V
18 22 V
18 21 V
18 22 V
17 21 V
18 21 V
18 22 V
18 21 V
18 22 V
18 21 V
18 21 V
18 22 V
17 21 V
18 22 V
18 21 V
18 21 V
18 22 V
18 21 V
18 22 V
18 21 V
17 21 V
18 22 V
18 21 V
18 22 V
18 21 V
18 21 V
18 22 V
18 21 V
18 22 V
17 21 V
18 21 V
18 22 V
18 21 V
18 22 V
18 21 V
18 21 V
18 22 V
17 21 V
18 22 V
stroke
grestore
end
showpage
}}%
\put(1989,423){\makebox(0,0)[r]{$\mathrm{Const}\cdot e^{{1\over2}\zeta_Nn}$}}%
\put(1989,542){\makebox(0,0)[r]{N=7}}%
\put(1989,661){\makebox(0,0)[r]{N=6}}%
\put(1989,780){\makebox(0,0)[r]{N=5}}%
\put(1989,899){\makebox(0,0)[r]{N=4}}%
\put(1989,1018){\makebox(0,0)[r]{N=3}}%
\put(1485,50){\makebox(0,0){n}}%
\put(200,1180){%
\special{ps: gsave currentpoint currentpoint translate
270 rotate neg exch neg exch translate}%
\makebox(0,0)[b]{\shortstack{$R_n/b_t$}}%
\special{ps: currentpoint grestore moveto}%
}%
\put(2308,200){\makebox(0,0){ 24}}%
\put(2171,200){\makebox(0,0){ 22}}%
\put(2034,200){\makebox(0,0){ 20}}%
\put(1897,200){\makebox(0,0){ 18}}%
\put(1759,200){\makebox(0,0){ 16}}%
\put(1622,200){\makebox(0,0){ 14}}%
\put(1485,200){\makebox(0,0){ 12}}%
\put(1348,200){\makebox(0,0){ 10}}%
\put(1211,200){\makebox(0,0){ 8}}%
\put(1073,200){\makebox(0,0){ 6}}%
\put(936,200){\makebox(0,0){ 4}}%
\put(799,200){\makebox(0,0){ 2}}%
\put(662,200){\makebox(0,0){ 0}}%
\put(550,562){\makebox(0,0)[r]{$10^{-3}$}}%
\put(550,936){\makebox(0,0)[r]{$10^{-2}$}}%
\put(550,1311){\makebox(0,0)[r]{$10^{-1}$}}%
\put(550,1685){\makebox(0,0)[r]{$1$}}%
\put(550,2060){\makebox(0,0)[r]{$10$}}%
\end{picture}%
\endgroup
 